# To Terminate or Not to Terminate Secure Sockets Layer (SSL) Traffic at the Load Balancer

Dr. Pierre D. Boisrond[i]

*e-Health Scholar / Computer Scientist / Social Scientist / Cybersecurity Researcher / Preacher*

*Abstract*

*The concepts of terminating or not terminating Secure Sockets Layer (SSL) at the load balancer have always generated intriguing conversations. In this paper, the author explains the pros and cons of such concepts in a simplistic manner and also provides suggested recommendations to help organizations understand the security implications associated with unencrypted traffic flowing from the Load Balancer to the App Servers.*

**OVERVIEW**

The process of Secure Sockets Layer (SSL) termination happens when the Load Balancer decrypts the SSL-encrypted data and sends it to the Application Server in an unencrypted manner (see figure 1). The advantage of such a technique is that the burden of performing any decryption is no longer the responsibility of the App Server, which in turn will increase the performance at the App Server level. In terms of Analyzing SSL Issues when dealing with various SSL attacks, it gives one a central location to focus during the investigation.

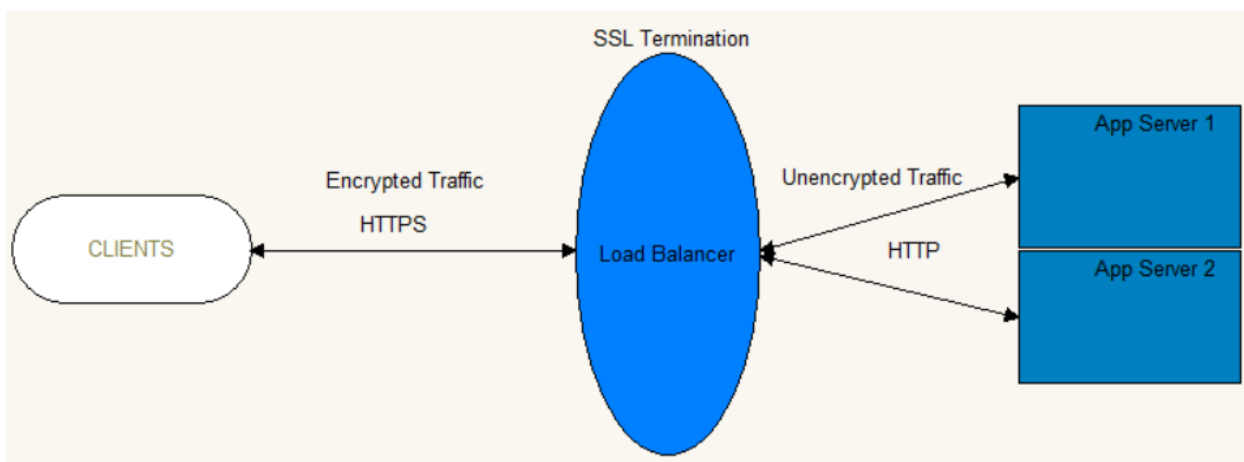

*Figure 1. SSL Traffic Termination at the Load Balancer*

However, such practice implies that SSL termination at the Load Balancer also poses a security risk because the data that are passing across the network from the load balancer to the App Server are now unencrypted, and that will leave applications vulnerable to Man-in-the-Middle Attack (MITM) (Boisrond, 2014). Having said that, one has to also take into consideration that the data in question are moving across the internal network, not the public Internet.  For example, if the deployment of the load balancer is behind a firewall on a secure subnet that only contains critical servers with no direct access to users, it might be acceptable to permit unencrypted data to pass from the Load Balancer to the App Server.  On the other hand, if the process of offloading the SSL to a firewall that resides on the edge of an organization's network, the risk of compromise is much greater after decrypting the data. Furthermore, when dealing with sensitive information, Payment Card Industry (PCI) compliance, and/or Health Insurance Portability and Accountability Act (HIPAA), the traffic between the load balancer and the App Servers must be re-encrypted due to government regulations. The same is true for traffic over an untrusted network.

**SUGGESTED RECOMMENDATIONS**

To address unencrypted traffic issues, organizations can do the following:

- Request an SSL certificate from a trusted Certificate Authority (CA) to be used between the Load Balancer and the App servers. That will give end-to-end encryption (E2EE).
- Since the decrypted traffic is in clear text, one can now place an Intrusion Detection System (IDS) / Intrusion Prevention System (IPS) between the Load Balancer and the App Servers or between the Load Balancer and the switch that the App servers are connected to protect them from Web Exploits. Access to the App Servers, the Load Balancers, and the IPS/IDS must be done via a secured management network.
- Use two Web Application Firewalls (WAF) to inspect the traffic. However, to keep the traffic encrypted, SSL certificates must be installed to encrypt the traffic from the Load Balancer to the WAF, and the traffic from the WAF to the App Servers ( See figure 2)

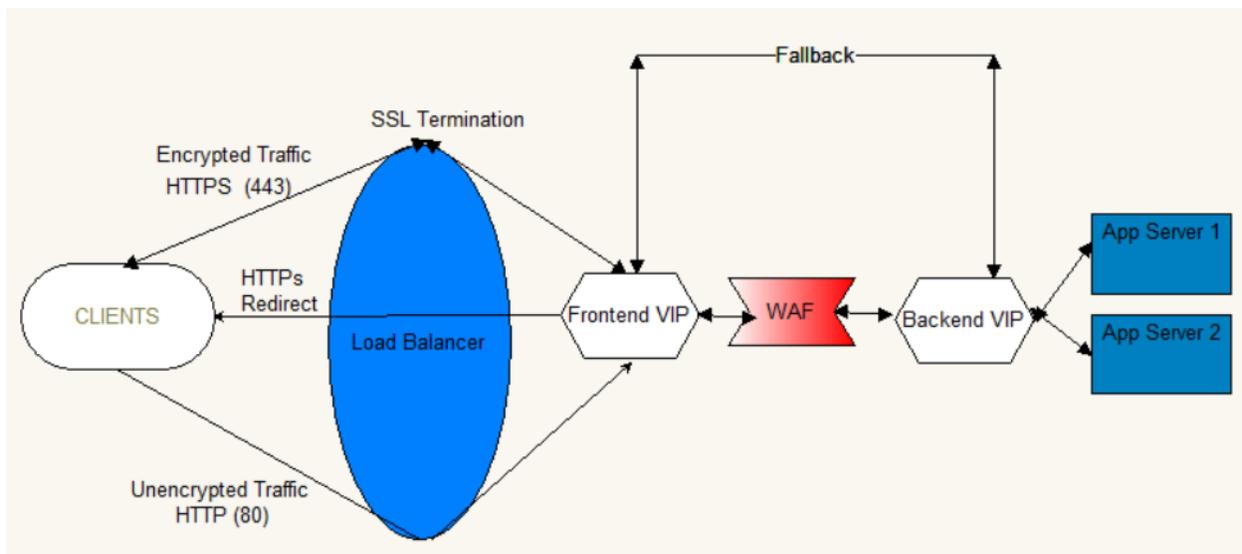

*Figure 2.  SSL Traffic Termination Implementation with Web Application Firewall (WAF)*

In regards to the above recommendations, if an organization does not have the proper protection in place to support them, the viable solution is to protect from end-to-end (E2EE) or accept the risk by requesting an exception to be evaluated by the Information Security Risk Evaluation Committee.

# CONCLUSION

There will always be a debate between security and performance when it comes to SSL encryption. However, organizations need to evaluate their network infrastructure, as well as the type of data on the network, and decide whether or not the extra security is worth the performance trade-off or vice-versa. Furthermore, one must take into consideration the expectations, as well as the perceptions of the customers. The reason is that many customers tend to assume that an encrypted connection is secured from E2EE, meaning from the client to the App Server. Therefore, an organization has to think about the legal consequences if confidential data of customers end up being breached or compromised/accessed by unauthorized personnel while traveling on the organization network in an unencrypted state or clear text.

---

[i] About the Author

*Dr. Pierre D. Boisrond is a cybersecurity researcher, a computer scientist, and an e-health scholar. He did his engineering and computer science research studies at Nova Southeastern University, FL; Cornell University, NY; and Boston University, MA. Regarding his research in health technology and medicine, Dr. Boisrond did his Health Informatics research studies at the University of Texas Health Science Center in Houston, TX, and at the College of Graduate Health Studies / Kirksville College of Osteopathic Medicine of A.T. Still University, MO., where he earned his doctorate.*

*Dr. Boisrond has also done medical informatics studies at the Feinberg School of Medicine at Northwestern University; Penn State College of Medicine; and The University of Medicine and Dentistry of New Jersey (UMDNJ), now Rutgers New Jersey Medical School. Dr. Boisrond was also accepted for graduate research studies at The Dartmouth Institute for Health Policy and Clinical Practice (TDI) at The Geisel School of Medicine of Dartmouth College in New Hampshire.*

*Dr. Boisrond is currently a reviewer for the Journal of Medical Internet Research and the Manager of Information Security and Cyber Threat at Sabre Corporation. Before Sabre Corporation, Dr. Boisrond frequented several prestigious engineering firms such as The Hewlett Packard Company, Raytheon Corporation, and the Boeing Company. Dr. Boisrond was sent to the University Hospital of Mirebalais in Haiti (https://www.pih.org/pages/mirebalais) on behalf of the Hewlett Packard Company and Partners in Health (https://www.pih.org/) to develop the information security policies for the hospital.*

*Dr. Boisrond currently lives in North Dallas with his wife, Kristen, and his two sons, Noah and Jacob*